\documentclass[aps,pra,showpacs,superscriptaddress,twocolumn,noeprint]{revtex4-1}

\usepackage{graphicx}
\usepackage{color}
\usepackage{amsmath}
\usepackage{float}
\usepackage{verbatim}
\usepackage{longtable}
\usepackage{multirow}
\usepackage{xcolor}
\usepackage{lineno}
\makeatletter
\newcommand{\rmnum}[1]{\it \romannumeral #1}
\newcommand{\Rmnum}[1]{\rm \expandafter\@slowromancap\romannumeral #1@}
\makeatletter 
\begin{document}
%%%%%%%%%%%%%%%%%%%%%%%%%%%%%%%%%%%%%%%%%%%%%%%%%%%%%%%%%%%%%%%%%%%%%%%%%%

\title{Rehabilitation of PBE-GGA for Layered Materials}

\author{Haowei Peng}
  	\email{Haowei.Peng@gmail.com. The implementation of $r$VV10 in VASP will be available within the next version.}
	\affiliation{Department of Physics, Temple University, Philadelphia, PA 19122, USA}
\author{John P. Perdew}
%	\email{perdew@temple.edu}
	\affiliation{Department of Physics, Temple University, Philadelphia, PA 19122, USA}
	\affiliation{Department of Chemistry, Temple University, Philadelphia, PA 19122, USA}
\begin{abstract}

The structural and energetic properties of layered materials propose a challenge to density functional theory with common semilocal approximations to the exchange-correlation. By combining the most-widely used semilocal generalized gradient approximation (GGA), Perdew--Burke--Ernzerhof (PBE), with the revised Vydrov--Van Voorhis non-local correlation functional (rVV10), both excellent structural and energetic properties of 28 layered materials were recovered with a judicious parameter selection. We term the resulting functional as PBE+rVV10L with ``L'' denoting for layered materials. Such combination is not new, and involves only refitting a single global parameter, however the resulting excellency suggests such corrected PBE for many aspects of theoretical studies on layered materials. For comparison, we also present the results for PBE+rVV10 where the parameter is determined by the 22 interaction energies between molecules.    

\end{abstract}


\maketitle

% document starts here!
% ====================================================

\section{Introduction}

Plenty of interests have been attracted by the two-dimensional (2D) materials and their parent layered materials \cite{Wang2012e, Chhowalla2013, Xu2013a, Butler2013a} since the experimentally realization of graphene \cite{Novoselov2004}. Layered materials presented a huge challenge for density functional theory (DFT) \cite{Kohn1965}, the current work-horse first-principles method. The difficulty comes from the coexistence of inter--layer van der Waals (vdW) interaction and strong chemical bonding within the layer. The vdW-correction needed for layered materials is usually weaker than that for molecular systems. Hence, most of vdW density functionals which are good for molecular systems \cite{Lee2010d, Cooper2010, Klimes2010a, Klimes2011, Hamada2014, Berland2014, Vydrov2009, Vydrov2010, Sabatini2013} overbind layered materials significantly \cite{Bjorkman2012a, Bjorkman2012, Bjorkman2012b, Bjorkman2014}. It is even not easy to find a dispersion-corrected generalized gradient approximation (GGA), which is able to predict well for both the geometric and energetic properties, i.e., the intra-layer lattice constant $a$, inter-layer lattice constant $c$, and inter-layer binding energy $E_b$ \cite{Bjorkman2012b, Bjorkman2014}. 

We have found a solution on the meta-GGA level, where we combined the strongly constrained appropriately normed (SCAN) \cite{Sun2015a} meta-GGA and the revised Vydrov--Van Voorhis non-local correlation functional (rVV10) \cite{Vydrov2010, Sabatini2013} with one parameter adjusted to the $Ar_2$ binding curve. The so-termed SCAN+rVV10 \cite{Peng2016} functional not only gives the best description for layered materials, but also excellently describes solids, molecular systems, adsorption of benzene on metal surfaces, and hence we expected it to be a versatile vdW density functional. One important conceptual feature of SCAN+rVV10, compared to other popular vdW density functionals \cite{Berland2015}, is that we deliberately combine the non-local vdW correlation functional with a semilocal functional which already includes certain amount of intermediate range vdW binding from the exchange. Instead, previous work was usually pursuing a vdW-free exchange functional.

Inspired by this new concept, we revisit the request of a GGA-based vdW density functional for layered materials, and end up with a solution by combining the most commonly used Perdew-Burke-Ernzerhof (PBE) \cite{Perdew1996} GGA and rVV10. The resulting functional, termed as PBE+rVV10L with the ``L'' denoting for layered materials, achieves similar accuracy as SCAN+rVV10. PBE+rVV10L is not as versatile as its meta-GGA counterpart SCAN+rVV10, however it is noticeably cheaper in computation, and numerically more stable thanks to the much simpler mathematical form of PBE. Besides, the rVV10 and PBE have been implemented in many ab-inito codes, and hence PBE+rVV10L provides a very handy solution for many problems related to layered materials. PBE+rVV10L is even better than the AM05-VV10sol functional \cite{Vydrov2010, Bjorkman2012b, Armiento2005}, which is constructed in a similar way as here but with an additional parameter adjusted (unfortunately a physically-sound justification has not been provided yet). Combining PBE and rVV10 (or VV10) is new in the condensed matter physics community, but is not in the quantum chemistry community where the PBE+VV10 has already been proposed for molecular systems \cite{Arago2013}. In this work, we will report the benchmarking results of this newly proposed PBE+rVV10L, and compared with the PBE+rVV10 where the parameter is adjusted to the interaction energies of 22 molecular complexes (S22) \cite{Podeszwa2010, Takatani2010} as the original VV10 \cite{Vydrov2010} and rVV10 \cite{Sabatini2013}. All calculations in this work were performed with the projector augmented wave (PAW) method \cite{Blochl1994b} as implemented in the VASP code (version 5.4.1) \cite{Kresse1994, Kresse1996, Kresse1999}. For more details, we refer to the Appendix in Ref.\,\cite{Peng2016}.

\begingroup
\squeezetable
\begin{table}[!htbp]                                                                                                   
\caption{Layer--layer binding energy $E_b$ in meV/\AA$^2$, inter--layer lattice constant $c$ in \AA, and intra--layer lattice constant $a$ in \AA\ for 28 layered materials from SCAN+rVV10 \cite{Peng2016} and PBE+rVV10L. The mean error (ME) and mean absolute error (MAE) are also given in the same units, and the mean relative error (MRE) and mean absolute relative errors (MARE) are given in percentage. The reference values for $E_b$ are from RPA calculations, and from experiments for $c$ and $a$ \cite{Bjorkman2012, Bjorkman2014}.}
\label{tab:2d}
\begin{ruledtabular}
\begin{tabular}{lccc|ccc|ccc}
  & \multicolumn{3}{c}{Reference}  & \multicolumn{3}{c}{SCAN+$r$VV10} & \multicolumn{3}{c}{PBE+$r$VV10L} \\
  &  $E_{b}$ & $c$  &  $a$ &  $E_{b}$ & $c$  &  $a$ &  $E_{b}$ & $c$  &  $a$   \\
\hline
TiS$_2$    & 18.88 & 5.71  & 3.41  & 18.90 & 5.68  & 3.40  & 18.04  &  5.79 & 3.39  \\
TiSe$_2$   & 17.39 & 6.00  & 3.54  & 18.53 & 6.02  & 3.54  & 18.99  &  6.07 & 3.52  \\
TiTe$_2$   & 19.76 & 6.50  & 3.78  & 19.74 & 6.59  & 3.75  & 22.65  &  6.55 & 3.75  \\
VS$_2$     & 25.61 & 5.75  & 3.22  & 20.67 & 5.81  & 3.17  & 20.20  &  5.92 & 3.17  \\
VSe$_2$    & 22.26 & 6.11  & 3.36  & 19.56 & 6.18  & 3.31  & 20.02  &  6.29 & 3.32  \\
VTe$_2$    & 20.39 & 6.58  & 3.64  & 19.69 & 6.84  & 3.54  & 22.59  &  6.74 & 3.58  \\
ZrS$_2$    & 16.98 & 5.81  & 3.66  & 15.95 & 5.79  & 3.67  & 15.12  &  5.93 & 3.66  \\
ZrSe$_2$   & 18.53 & 6.13  & 3.77  & 16.54 & 6.12  & 3.78  & 16.32  &  6.24 & 3.77  \\
ZrTe$_2$   & 16.34 & 6.66  & 3.95  & 19.53 & 6.67  & 3.97  & 21.15  &  6.69 & 3.93  \\
NbS$_2$    & 17.58 & 17.91 & 3.33  & 20.20 & 18.11 & 3.33  & 19.78  & 18.42 & 3.33  \\
NbSe$_2$   & 19.57 & 12.55 & 3.44  & 21.37 & 12.55 & 3.45  & 21.96  & 12.65 & 3.46  \\
NbTe$_2$   & 23.03 & 6.61  & 3.68  & 21.83 & 6.88  & 3.64  & 23.51  &  6.84 & 3.67  \\
MoS$_2$    & 20.53 & 12.30 & 3.16  & 19.89 & 12.28 & 3.16  & 19.24  & 12.57 & 3.17  \\
MoSe$_2$   & 19.63 & 12.93 & 3.29  & 19.33 & 13.01 & 3.29  & 19.25  & 13.23 & 3.31  \\
MoTe$_2$   & 20.80 & 13.97 & 3.52  & 20.45 & 14.14 & 3.50  & 21.40  & 14.13 & 3.53  \\
PdTe$_2$   & 40.17 & 5.11  & 4.02  & 41.74 & 5.00  & 4.03  & 41.71  &  5.13 & 4.08  \\
HfS$_2$    & 16.13 & 5.84  & 3.63  & 15.85 & 5.79  & 3.61  & 15.05  &  5.97 & 3.62  \\
HfSe$_2$   & 17.09 & 6.16  & 3.75  & 16.10 & 6.14  & 3.73  & 15.80  &  6.27 & 3.74  \\
HfTe$_2$   & 18.68 & 6.65  & 3.96  & 17.99 & 6.69  & 3.94  & 19.36  &  6.73 & 3.93  \\
TaS$_2$    & 17.68 & 5.90  & 3.36  & 21.11 & 5.88  & 3.35  & 20.32  &  6.03 & 3.35  \\
TaSe$_2$   & 19.44 & 6.27  & 3.48  & 20.67 & 6.27  & 3.46  & 20.82  &  6.38 & 3.47  \\
WS$_2$     & 20.24 & 12.32 & 3.15  & 20.08 & 12.35 & 3.15  & 19.59  & 12.68 & 3.17  \\
WSe$_2$    & 19.98 & 12.96 & 3.28  & 19.82 & 13.03 & 3.28  & 19.72  & 13.28 & 3.30  \\
PtS$_2$    & 20.55 & 5.04  & 3.54  & 18.82 & 5.06  & 3.53  & 18.01  &  5.09 & 3.58  \\
PtSe$_2$   & 19.05 & 5.08  & 3.73  & 19.02 & 5.25  & 3.71  & 19.79  &  5.01 & 3.79  \\
Gra.   & 18.32 & 6.70  & 2.46  & 20.30 & 6.54  & 2.45  & 16.04  &  6.90 & 2.47  \\
$h$-BN     & 14.49 & 6.69  & 2.51  & 18.45 & 6.48  & 2.50  & 14.43  &  6.85 & 2.51  \\
PbO        & 20.25 & 5.00  & 3.96  & 22.93 & 4.81  & 3.98  & 17.95  &  5.08 & 4.04  \\
ME   & & & & 0.20 & 0.02 & -0.01 &-0.02 & 0.15 & 0.00 \\
MAE	 & & & & 1.48 & 0.08 & 0.02 & 1.74 & 0.15 & 0.02 \\
MRE  & & & & 1.7  & 0.2 & -0.4  & 0.2  & 1.8  & 0.0 \\ 
MARE & & & & 7.7  & 1.2  & 0.5  & 8.9  & 1.9  & 0.7 \\
\end{tabular}
\end{ruledtabular}
\end{table}
\endgroup

The $r$VV10 \cite{Vydrov2010, Sabatini2013} nonlocal correlation functional takes a similar form as the popular family of Rutgers-Chalmers vdW-DFs \cite{Lee2010d, Cooper2010, Klimes2010a, Klimes2011, Hamada2014, Berland2014},
\begin{equation}
\label{eq:ec_nl}
E_c^{nl} = \int\!\!d\boldsymbol{r} n(\boldsymbol{r}) [\frac{\hbar}{2}\int\!\!d\boldsymbol{r^{\prime}}  \Phi(\boldsymbol{r}, \boldsymbol{r^{\prime}}) n(\boldsymbol{r^{\prime}})+\beta],
\end{equation}
$\beta$ vanishes for the Rutgers-Chalmers vdW-DFs, and the total exchange correlation functional reads
\begin{equation}
\label{eq:ex}
E_{xc} = E_{xc}^0 + E_c^{nl}.
\end{equation}
Here $n(\boldsymbol{r})$ is the electron density, and $\Phi(\boldsymbol{r}, \boldsymbol{r^{\prime}})$ is the kernel describing the density-density interactions, $E_{xc}^0$ is the companying semilocal exchange correlation. To ensure zero $E_c^{nl}$ for the uniform electron gas, $\beta=\frac{1}{32}(\frac{3}{b})^{\frac{3}{4}}$ in Hartree is required. Two empirical parameters $C$ and $b$ appear in the kernel $\Phi(\boldsymbol{r}, \boldsymbol{r^{\prime}})$: $C$ chosen for accurate $-C_6/R^6$ vdW interactions between molecules at large separation $R$, and $b$ controlling the damping of $E_c^{nl}$ at short range. 

In the original form for both VV10 and rVV10 \cite{Vydrov2010, Sabatini2013}, the nonlocal correlation was proposed to combine with the semilocal exchange-correlation functional \cite{Murray2009a, Perdew1996} $E_{xc}^0 = E_x^{rPW86}+E_c^{PBE}$, partly because of the rPW86 exchange is nearly vdW-free \cite{Murray2009a}. For a semilocal $E_{xc}^0$, $C=0.0093$ was generally recommended \cite{Vydrov2010}, and the $b$ parameter was determined as $5.9$ and $6.3$ by fitting to the interaction energies of the S22 set \cite{Podeszwa2010, Takatani2010} for the original VV10 and $r$VV10. Increasing $C$ or $b$ generally results in smaller vdW correction. Keeping the original semilocal part, $b=9.15$ is required to fit the binding energies of 26 layered materials for both VV10 \cite{Bjorkman2012b} and rVV10 \cite{Peng2016}, implying that layered materials require weaker vdW correction than molecular complexes. However such fitted $b$ value leads to much worse performance for both the intra- and inter-layer lattice constants in layered materials, and also for solids \cite{Bjorkman2012b}. Besides, $b=9.3$ in rVV10 was proposed for the structural properties of water \cite{Miceli2015} where hydrogen bonding matters more, since the original VV10 and rVV10 overbind the seven hydrogen-bonded complexes from the S22 \cite{Vydrov2010, Sabatini2013}.

Changing the semilocal $E_{xc}^0$ to the SCAN meta-GGA \cite{Sun2015a} results in the versatile SCAN+rVV10 with $b=15.7$ \cite{Peng2016}, and the SCAN+VV10 with $b=14.1$ \cite{Brandenburg2016b}. Selecting $E_{xc}^0$ to the AM05 form \cite{Armiento2005} results in the AM05-VV10sol with $b=10.25$ and $C=10^{-6}$ \cite{Bjorkman2012b}, which works well for layered materials but not for S22. Note the practically zero value for the $C$ parameter is required by the fitting but not by chemistry. For PBE+VV10, $b=6.2$ is determined by fit to S22 \citep{Arago2013}. In this work, we determine $b=10.0$ for PBE+rVV10L by fitting to the inter-layer binding energies of 28 layered materials \cite{Bjorkman2012, Bjorkman2012b, Bjorkman2014, Peng2016}, and $b=6.6$ for PBE+rVV10 by fitting to S22. The $b$ value for PBE+rVV10 is slightly larger than that for the original rVV10, in accordance with the slightly more vdW binding from the PBE exchange than that from the rPW86 exchange \cite{Murray2009a}. This difference is related to that the exchange enhancement factor of PBE is bounded by the Lieb-Oxford constraintas as the reduced density gradient $s$ increases \cite{Perdew1996}, while the rPW86 enhancement factor diverges as $s^{0.4}$ \cite{Murray2009a}. The reference binding energies of the layered materials are not from experiments, but from adiabatic-connection fluctuation-dissipation theorem within the random-phase approximation (RPA) \cite{Harl2009, Harl2010, Eshuis2012}, which are yet the best available choice. 

%(In the following, we use ``$r$VV10'' to specifically denote the original $r$VV10 density functional with $E_{xc}^0 = E_x^{rPW86}+E_c^{PBE}$, $C=0.0093$, and $b=6.3$ \cite{Sabatini2013}.)

The lattice constants, both intra-layer $a$ and inter-layer $c$, and the layer--layer binding energy $E_b$ are the most fundamental quantities when one embarks on first-principles computation of layered materials. For the benchmarking, we use the binding energies from RPA, and lattice constants from experiments as references \cite{Bjorkman2012, Bjorkman2014}. Until now, SCAN+rVV10 is the only one which can simultaneously predict with the mean absolute relative error $<10\%$ for $E_b$, $<2\%$ for $c$, and $<1\%$ for $a$ \cite{Peng2016}. In Table\,\ref{tab:2d}, we compare the results from PBE+rVV10L to those from SCAN+rVV10, and the reference values. PBE+rVV10L achieves excellent accuracy for both the geometrical and energetic properties by adjusting only one parameter, which is not trivial as discussed above. PBE+rVV10L actually is comparable with SCAN+rVV10 for this class of materials, with a slightly overestimated layer-layer spacing. Nevertheless, PBE+rVV10L is another member of the ``10-2-1'' club for layered materials. Considering the extremely simply mathematical form of both PBE and rVV10, and their widely availability in many scientific codes, PBE+rVV10L can be a very handy theoretical tool for layered materials. It can be also used to prepare reasonably good initial (relaxed) structure and orbitals to accelerate the convergence of following SCAN+rVV10 calculations. 

\begin{figure}[!htbp]                                                                                                   
\begin{center}
\caption{(Color online) Box-plots for the absolute relative errors and relative errors of the inter-layer binding energies, inter- and intra-layer lattice constants ($c$ and $a$) from SCAN+rVV10 \cite{Peng2016}, PBE+rVV10, and PBE+rVV10L, for 28 layered materials. The reference values are from RPA for the binding energy, and from experiment for the lattice constants \cite{Bjorkman2012, Bjorkman2014}. The Tukey box-plot used here summarizes the overall distribution of a set of data points: The bottom and the top of the box are the first (Q1) and third (Q3) quartiles (25\% of data points lies below Q1, and another 25\% above Q3); The band inside the box denotes the median; The circles if any denote outliers which lie further than 1.5$*|Q3-Q1|$ away from the box; The vertical line extends from the minimum to the maximum, except for the outliers. Besides, we also denote the mean value with a shape inside the box.}
\label{fig:2D}
\includegraphics[width=3.4in]{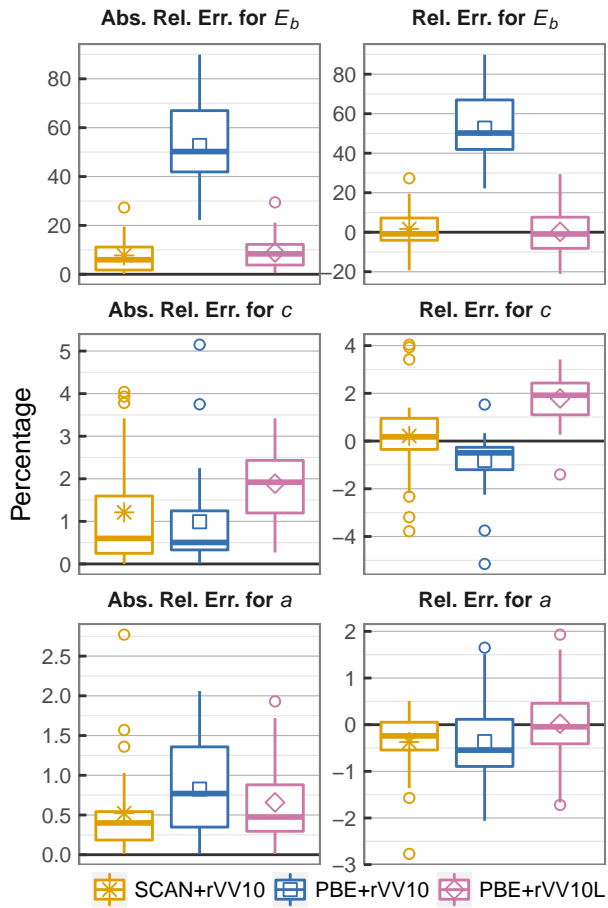}
\end{center}
\end{figure}

In Fig.\,\ref{fig:2D}, we summarise the absolute relative errors and relative errors for $a$, $c$ and $E_b$ from $SCAN+rVV10$, $PBE+rVV10$, and $PBE+rVV10L$. With a smaller $b$ parameter, PBE+rVV10 overbinds the layered materials by about 50\%, similar to the original rVV10, but the lattice constants are still reasonably accurate. Therefore, a $b$ parameter between $6.6$ and $10.0$ may be empirically chosen in case that the accuracy for the layer--layer binding energy could be less relevant.

\begin{figure}[!htbp]                                                                                                   
\begin{center}
\caption{(Color online) Box-plots for the absolute relative errors and relative errors of the atomization energies $E_a$, and the lattice equilibrium volumes $V_0$, from RPA, PBE, PBE+rVV10, and PBE+$r$VV10L for 50 solids, with respect to the experimental values. The RPA, PBE, and experimental values (after the zero-point correction) are from Refs.\,\onlinecite{Harl2010a} and \onlinecite{Schimka2013a}. }
\label{fig:solid}
\includegraphics[width=3.4in]{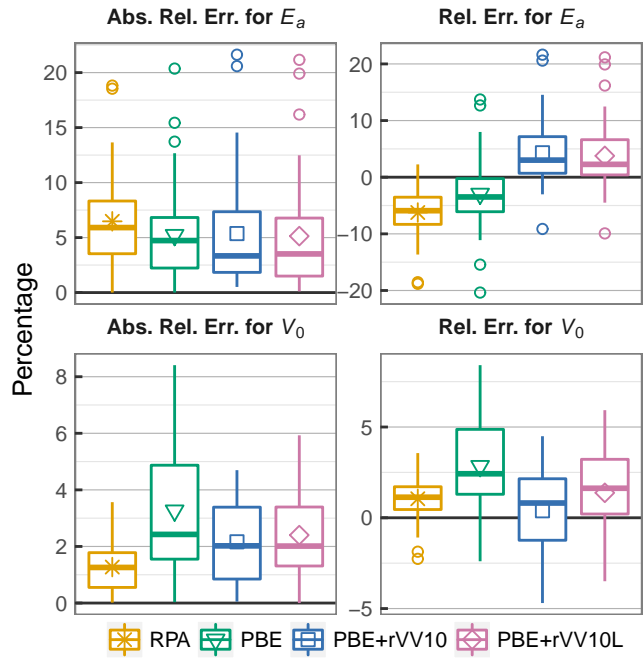}
\end{center}
\end{figure}

For solid systems, we further benchmark the performance of PBE+rVV10L and PBE+rVV10 for 50 solids compiled as in Ref.\, \onlinecite{Peng2016}, which includes ($\rmnum{1}$) 13 group--$\Rmnum{4}$ and $\Rmnum{3}$--$\Rmnum{5}$ semiconductors, ($\rmnum{2}$) 5 insulators, ($\rmnum{3}$) 8 main-group metals, ($\rmnum{4}$) 3 ferromagnetic transition metals Fe, Co, and Ni, and ($\rmnum{5}$) 21 other transition metals for which non-spin-polarized calculations were performed \cite{Harl2010a, Schimka2013a}. We compared the mean relative errors and mean absolute errors for atomization energies and lattice volumes. The rVV10 correction decreases the mean absolute relative error for the atomization energies slightly for PBE, and the otherwise underestimated atomization energies are slightly overestimated now with both PBE+rVV10 and PBE+rVV10L. However, atomization energy may not be a good choice to assess a semilocal functional \cite{Perdew2015}. It is well-known that PBE overestimate the lattice volume, with a mean absolute relative error over 3\% as shown in Fig.\, \ref{fig:solid}. The attractive vdW correction slightly remedies this systematic overestimation by about 1\%. This has been already known in recent works of Tao et al.\,\cite{Tao2010, Tao2016z}. Overall, the structure and energetic properties for solids are not skewed by the rVV10 correction with both PBE+rVV10 and PBE+rVV10L.

\begin{figure}[!htbp]                                                                                                   
\begin{center}
\caption{(Color online) Box-plots for the absolute relative errors and relative errors of the interaction energies from $r$VV10, SCAN+rVV10, PBE+rVV10, PBE+rVV10L, with respect to the CCSD(T) results \cite{Podeszwa2010, Takatani2010}, for the molecular complexes in the S22 dataset.}
\label{fig:S22}
\includegraphics[width=3.4in]{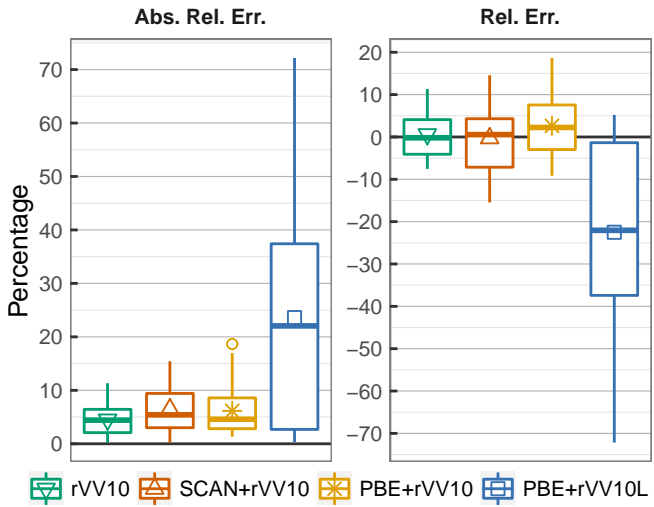}
\end{center}
\end{figure}

Fig.\,\ref{fig:S22} shows the results of PBE+rVV10 and PBE+rVV10L for S22 which includes seven hydrogen-bonded, eight dispersion-bound, and seven mixed complexes, compared to the rVV10, SCAN+rVV10, and the CCSD(T) reference \cite{Podeszwa2010, Takatani2010}. Similar to PBE+VV10 \cite{Arago2013}, the fitting of PBE+rVV10 is less accurate than the original rVV10 with the rPW86 exchange, and the mean absolute relative error of 6\% (4.5\% for rVV10). PBE+rVV10L with the $b$ parameter fitted to layered materials significantly underbinds with a mean absolute relative error of 24\% and a mean relative error of 22\%. Nevertheless, PBE+rVV10L is noticeably better than AM05-VV10sol, whose mean absolute relative error is 36\% \cite{Bjorkman2012b}. Besides, PBE+rVV10L performs very well for the seven hydrogen-bonding complexes with a mean absolute relative error of only $2\%$. This indicates that PBE+rVV10L should be better than PBE+rVV10 for structural properties of water \cite{Miceli2015}.

\begin{table}[!htbp]
\caption{Adsorption energy $E_{ad}$ and distance $\Delta_z$ between benzene and the (111) surface of Cu, Ag, and Au from PBE+rVV10 and PBE+$r$VV10L, compared with the SCAN+rVV10 results \cite{Peng2016}. The data for the lowest-energy hcp30$^{\circ}$ configuration \cite{Liu2013x} is shown.}
\label{tab:bz}
\begin{ruledtabular}
\begin{tabular}{lcccccc}
	& \multicolumn{2}{c}{PBE+rVV10}  &  \multicolumn{2}{c}{PBE+$r$VV10L} & \multicolumn{2}{c}{SCAN+rVV10}\\
\hline
	& $E_{ad}$ (eV) & $\Delta_z$ (\AA) & $E_{ad}$ (eV) & $\Delta_z$ (\AA) & $E_{ad}$ (eV) & $\Delta_z$ (\AA) \\
Cu	&	$0.84$ & $2.88$ 	& $0.52$ & $3.05$ & $0.74$ & $2.93$ \\
Ag 	& 	$0.74$ & $3.02$	& $0.46$ & $3.18$ & $0.68$ & $3.02$ \\
Au	& 	$0.82$ & $3.04$	& $0.51$ & $3.20$ & $0.73$ & $3.07$ \\
\end{tabular}
\end{ruledtabular}
\end{table}

At last, we benchmarked the performance of PBE+rVV10 and PBE+rVV10L with the adsorption of Benzene ring on Cu, Ag and Au (111) surfaces, which have been widely studied \cite{Bilic2006a, Toyoda2009, Liu2013x, Yildirim2013, Reckien2014, Carter2014}. The SCAN+rVV10 results \cite{Peng2016} are chosen for reference, which agrees very well with available experiments \cite{Liu2015j, Campbell2012, Xi1994, Zhou1990, Syomin2001, Ferrighi2011a}. In these systems, PBE+rVV10L is slightly worse than PBE+rVV10, underestimating the binding energy $\Delta E$ by about 0.2\,eV and overestimating the distance $\Delta_z$ between benzene and metal surface by about 0.14\,\AA. PBE+rVV10 is better than PBE+rVV10L, and only overbinds comparing to SCAN+rVV10 very slightly.   

In conclusion, we provide here two set of parameters for the combination between PBE and rVV10. For systems involving molecules, the PBE+rVV10, where $b=6.6$, gives better prediction. For layered materials (and perhaps also hydrogen-bonding systems), PBE+rVV10L, where $b=10.0$, achieves the accuracy of the best dispersion-corrected, semilocal density functional, and hence is highly recommended. Values between these two may also be employed for specific problems. The PBE+VV10L and PBE+VV10 is not as versatile as the meta-GGA-level SCAN+rVV10, but they are very handy and computationally high-efficiency alternatives. 

\bibliography{/home/haowei/papers/000000_BibTeX/references}
\end{document}